%% file: EXO-10-002_temp.tex
\pdfoutput=1
\documentclass[11pt,twoside,a4paper,cmspaper,final,collab]{cms-tdr}
\def\svnVersion{18891M}\def\cmsCernNo{2010-038}\def\cmsCernDate{2010/10/21}\def\cmsMessage{Submitted to Physical Review Letters}

\begin{document}\cmsNoteHeader{EXO-10-002}
%
%
%

%
%
\hyphenation{env-iron-men-tal}
\hyphenation{had-ron-i-za-tion}
\hyphenation{cal-or-i-me-ter}
\hyphenation{de-vices}
%
\RCS$Revision: 18842 $
\RCS$HeadURL: svn+ssh://alverson@svn.cern.ch/reps/tdr2/papers/EXO-10-002/trunk/EXO-10-002.tex $
\RCS$Id: EXO-10-002.tex 18842 2010-10-20 12:07:16Z jhirsch $

\input{ptdr-definitions}

\cmsNoteHeader{EXO-10-002} 
\title{Search for Quark Compositeness with the Dijet Centrality Ratio in pp Collisions at $\sqrt{s}=7$ TeV}

\newcommand {\PYTHIABARE}{{\textsc{pythia}}\xspace}
\newcommand {\NLOJET}{{\textsc{nlojet++}}\xspace}
\newcommand {\fastNLO}{{\textsc{fastNLO}}\xspace}
\renewcommand {\HERWIG}{{\textsc{herwig++}}\xspace}
\renewcommand {\PYTHIA}{{\textsc{pythia6}}\xspace}


\date{\today}

\abstract{
A search for quark compositeness in the form of quark contact interactions,
based on hadronic jet pairs (dijets) produced in proton-proton collisions at
$\sqrt{s}=7$ TeV, is described.  The data sample of the study corresponds to
an integrated luminosity of 2.9 pb$^{-1}$ collected with the CMS detector at
the LHC.  The dijet centrality ratio, which quantifies the angular
distribution of the dijets, is measured as a function of the invariant mass
of the dijet system and is found to agree with the predictions of the
Standard Model.  A statistical analysis of the data provides a lower limit
on the energy scale of quark contact interactions.  The sensitivity of the
analysis is such that the expected limit is 2.9 TeV; because the observed
value of the centrality ratio at high invariant mass is below the
expectation, the observed limit is 4.0 TeV at the 95\% confidence level.}



\hypersetup{
pdfauthor={CMS Collaboration},
pdftitle={Search for Quark Compositeness with the Dijet Centrality Ratio in pp Collisions at $\sqrt{s}=7$ TeV},%
pdfsubject={CMS},%
pdfkeywords={CMS, physics, software, computing}}

\maketitle 




In the Standard Model (SM), most high energy proton-proton collisions
are described by the scattering of partons (quarks or gluons) in the
framework of Quantum Chromodynamics (QCD).  The outgoing partons manifest
themselves as two or more jets of hadrons, with the pseudorapidity $\eta$ of
the jets depending on the parton scattering angle.  In QCD, the jet
production rate peaks at large $\vert\eta\vert$ because the scattering is
dominated by $t$-channel processes.  Several new physics scenarios,
including models of quark compositeness, produce a more isotropic angular
distribution leading to enhanced jet production at smaller values of
$\vert\eta\vert$~\cite{ELP,ref_qstar,ref_diquark,ref_axi,ref_coloron,
ref_rsg}.  Other models of new physics predict the opposite: a decrease in
jet production at small $\vert\eta\vert$ compared with the SM~\cite{meade}.

The dijet system consists of the two jets with the highest transverse
momenta $p_T$ in an event (the leading jets) with invariant mass $m_{jj}$.
The dijet centrality ratio $R_{\eta}$ is defined as the number of events
with the two leading jets in the region $|\eta|<0.7$ (\textit{inner} events)
divided by the number of events with the two jets in the region
$0.7<|\eta|<1.3$ (\textit{outer} events).  Because many systematic effects
cancel in this ratio, $R_{\eta}$ provides an accurate test of QCD and is
sensitive to new physics.

In this Letter we report a measurement of $R_{\eta}$ in proton-proton
collisions at $\sqrt{s}=7$ TeV.  The analysis is based on a data sample
corresponding to an integrated luminosity of $2.9 \pm 0.3$ pb$^{-1}$
collected with the Compact Muon Solenoid (CMS) detector at the CERN Large
Hadron Collider.

The dijet centrality ratio is measured as a function of $m_{jj}$ and is
compared with predictions from perturbative QCD calculations performed at
next-to-leading order (NLO) accuracy with the \NLOJET
program~\cite{nlo1,nlo2} in the \fastNLO framework~\cite{fastnlo}.  We also
compare the measured $R_{\eta}$ with a QCD prediction obtained from the
\PYTHIABARE 6.420 event generator~\cite{pythia} with the D6T set of
parameters~\cite{d6t}.  We use CTEQ6.6 parton distribution functions
(PDFs)~\cite{cteq} for the NLO calculation and CTEQ6LL~\cite{cteq6} for the
\PYTHIA prediction.  The effect of the CMS detector simulation~\cite{geant}
on the predictions for $R_{\eta}$ is negligible.

We use $R_{\eta}$ to search for evidence that quarks are composite
particles.  Quark compositeness at an energy scale $\Lambda$ would appear at
lower energies as a contact interaction, yielding an $\eta$ distribution
different from that predicted by QCD.  We consider a model of contact
interactions between left-handed quarks in the process $qq\rightarrow qq$
described by the effective Lagrangian $L_{qq} = (\pm 2\pi/\Lambda^{2})
(\overline{q}_{L}\gamma^{\mu}q_L) (\overline{q}_L \gamma_\mu
q_L)$~\cite{ELP}.  We choose the positive sign of $L_{qq}$ because it yields
a more conservative exclusion limit on $\Lambda$ (by about $5\%$) than the
negative sign.  In QCD, $R_{\eta}$ is nearly independent of $m_{jj}$, with a
value near $0.5$.  The presence of quark contact interactions described by
$L_{qq}$ would cause $R_{\eta}$ to increase rapidly above a value of
$m_{jj}$ that depends on $\Lambda$.  Previous searches for quark
compositeness described by the interaction $L_{qq}$ exclude $\Lambda<3.4$
TeV at the $95\%$ confidence level
(CL)~\cite{ua2,ua1,cdf,dzero_rat,dzero_ang,atlas}.

A detailed description of the CMS detector can be found
elsewhere~\cite{cmsdet}.  The CMS coordinate system has the origin at the
center of the detector, the $z$-axis along the direction of the
counterclockwise beam, and the transverse plane perpendicular to the beam
axis; $\phi$ is the azimuthal angle, $\theta$ the polar angle, and $\eta
\equiv -\ln(\tan[\theta/2])$ the pseudorapidity.  The central feature of the
CMS apparatus is a superconducting solenoid that surrounds the silicon pixel
and strip tracker as well as the barrel and endcap calorimeters (covering
the region $|\eta|<3$): a lead tungstate crystal electromagnetic calorimeter
(ECAL) and a brass-scintillator hadronic calorimeter (HCAL).  The ECAL
barrel extends to $\vert\eta\vert=1.479$ and the HCAL barrel to
$\vert\eta\vert=1.305$.  The HCAL and ECAL cells are grouped into towers
projecting radially outward from the origin.  In the region
$\vert\eta\vert<1.74$ these calorimeter towers have width $\Delta\eta =
\Delta\phi = 0.087$.  ECAL and HCAL cell energies above noise suppression
thresholds are summed within each projective tower to define the calorimeter
tower energy.

We reconstruct jets by applying the anti-$k_T$ clustering
algorithm~\cite{antikt} to the calorimeter towers with the
distance parameter $R=0.7$.  The energy $E$ and momentum $\vec{p}$ of a jet are
defined as the scalar and vector sums, respectively, of the calorimeter cell
energies associated with the jet.  We apply $p_T$- and $\eta$-dependent
scales to $E$ and $\vec{p}$ to correct for the non-linearity and
non-uniformity of the calorimeter response.  The jet energy corrections and
resolutions are determined and validated using simulated, test beam, and
collision data~\cite{jetPAS}.

A set of independent single-jet triggers with varying thresholds on
uncorrected jet $p_T$ is employed in the online trigger system. We use data
from three of these triggers, with thresholds of $15$, $30$, and $50$ GeV,
in the $m_{jj}$ ranges where the triggers have efficiency greater than
$99.5\%$ for both inner and outer events.  By studying the relative
efficiency of parallel triggers in the collision data, we determine that
these three triggers meet this efficiency requirement for $m_{jj}$ greater
than $156$, $244$, and $354$ GeV, respectively, where these values are three
of the predefined bin edges for $m_{jj}$.  The requirement of $m_{jj}>156$
GeV results in a minimum jet $p_T$ of $25$ GeV.

To remove potential instrumental and non-collision backgrounds we impose the
following requirements: events must have a primary vertex reconstructed with
$\vert z\vert<24$ cm~\cite{trkpas}; jets must have at least $1\%$ of their
total energy detected in the ECAL, no more than $98\%$ of their energy
detected in a single HCAL photodetector, and no more than $90\%$ of their
energy in a single calorimeter cell (ECAL or HCAL).  These jet
identification criteria remove less than $0.1\%$ of the events that pass the
$m_{jj}$ threshold and $\eta$ constraints.

\begin{figure}[hbtp!]
  \begin{center}
  \includegraphics[scale=0.39]{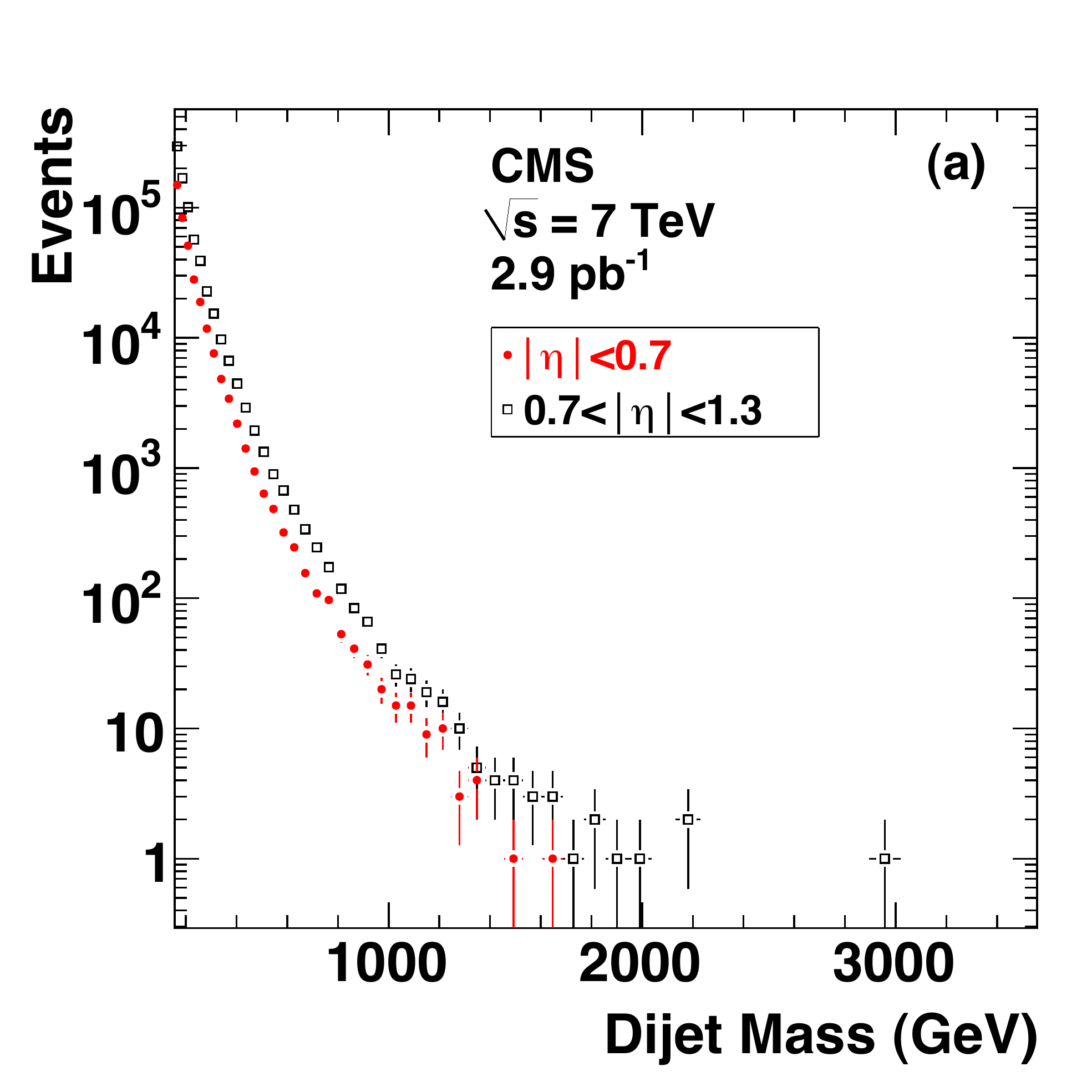}
  \includegraphics[scale=0.39]{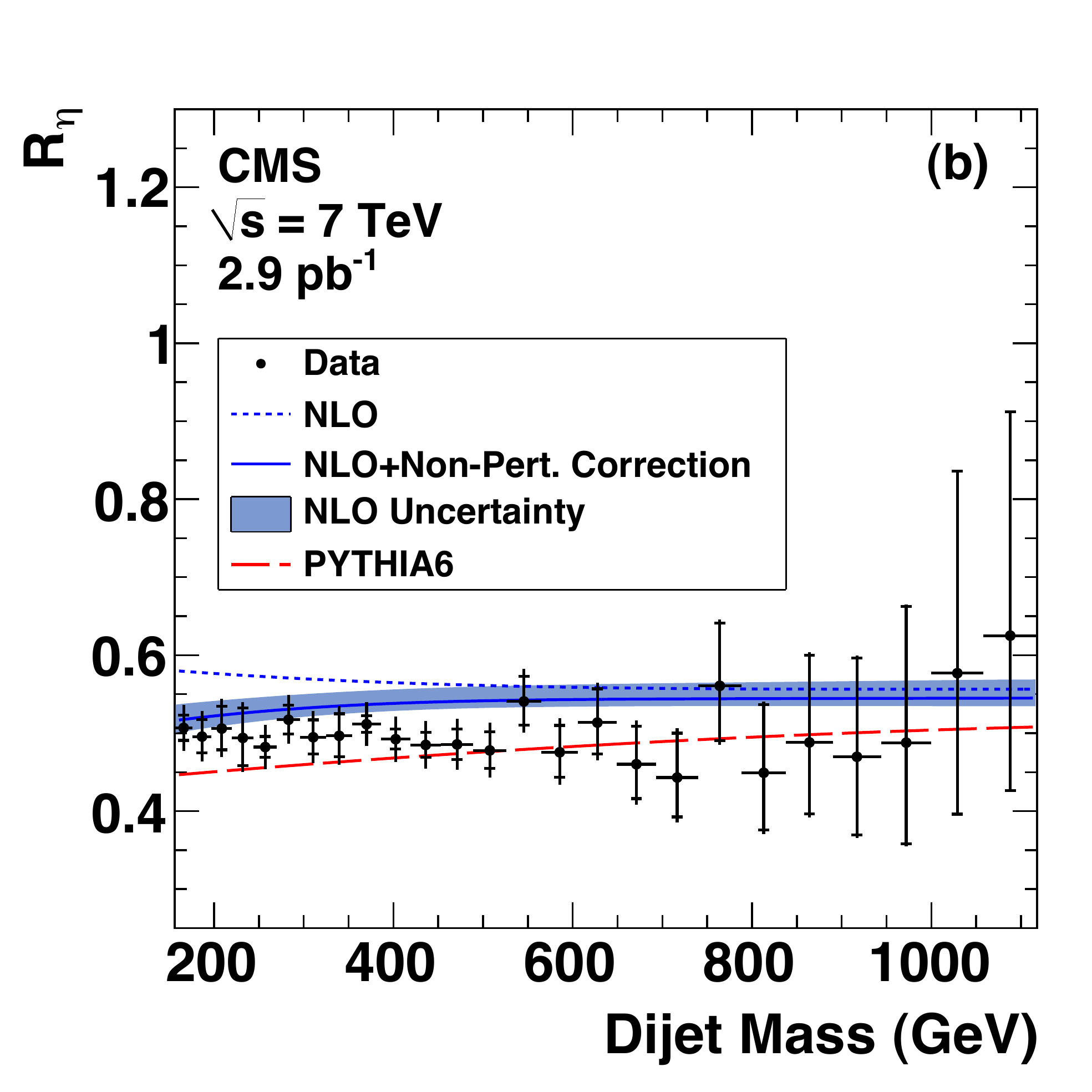}
\caption{(a) Event counts corrected for trigger prescales for
  inner (solid circles) and outer (open boxes) dijets and (b) the observed
  $R_{\eta}$ as functions of $m_{jj}$.  We compare $R_{\eta}$ with
  predictions for QCD from \PYTHIA (dashed line), NLO
  calculations (dotted line), and NLO plus non-perturbative
  corrections (solid line) and its uncertainty (band).}
\label{fig:data}
\end{center}
\end{figure}

In Fig.~\ref{fig:data} we show the observed numbers of inner and outer dijet
events and $R_{\eta}$ in bins of $m_{jj}$; the bin widths roughly correspond
to the $m_{jj}$ resolution.  The event counts, which are corrected for the
trigger reduction factors (prescales), fall steeply with increasing
$m_{jj}$.  We compare $R_{\eta}$ with NLO and \PYTHIA predictions for
$m_{jj}$ values up to $1120$ GeV.  The error bars represent the combination
of statistical and experimental systematic uncertainties (described in
detail later).  The horizontal lines near the end of the error bars denote
the statistical uncertainty on this ratio of Poisson-distributed variables
computed with Clopper-Pearson intervals~\cite{ClopPear}.

We apply an $m_{jj}$-dependent correction to the NLO prediction to account
for non-perturbative effects of hadronization and multiple parton
interactions.  This correction, which is approximately $10\%$ at low
$m_{jj}$ and $2\%$ for $m_{jj}$ greater than 400 GeV, is obtained from
\PYTHIA.  The predictions of \PYTHIA and \HERWIG~\cite{herwig} for this
correction agree to within a few percent.

The NLO prediction is shown as a band that accounts for uncertainties
related to the choices of the renormalization scale $\mu_{\rm R}$, the
factorization scale $\mu_{\rm F}$, and the PDFs used in the calculation.
The scale uncertainties, which are approximately 3--4\% depending on
$m_{jj}$, are evaluated by varying the scales from the default choice of
$\mu_{\rm R} = \mu_{\rm F} = p_T$ to $p_T/2$, $p_T$, and $2p_T$ in the
following six combinations: $(\mu_{\rm R},\mu_{\rm F}) = (p_T/2, p_T/2)$,
$(2p_T, 2p_T)$, $(p_T, p_T/2)$, $(p_T, 2p_T)$, $(p_T/2, p_T)$, and $(2p_T,
p_T)$.  The PDF uncertainties are estimated with repeated evaluations of the
NLO-predicted $R_{\eta}$ for the PDFs in the CTEQ6.6, MSTW2008~\cite{mstw},
and NNPDF2.0~\cite{nnpdf} sets and are found to be less than $1\%$.  The
band also includes the uncertainty arising from the correction for
non-perturbative effects, which we conservatively take to be $20\%$ of the
correction factor.


The measured $R_\eta$ is nearly flat with a value of about $0.5$ as
predicted by both the corrected NLO calculation and \PYTHIA.  The observed
average ratio is about $7\%$ lower than that of the corrected NLO
prediction, and about $7\%$ higher than that of the \PYTHIA prediction.  The
data are in better agreement with the corrected NLO prediction at low
$m_{jj}$, where the significant non-perturbative corrections improve the
agreement, and with the \PYTHIA prediction at intermediate and high
$m_{jj}$.

To test for the presence of quark compositeness with $R_{\eta}$, we employ a
log-likelihood-ratio statistic that compares the likelihood of the null (QCD
only) hypothesis ${\cal L}_{\rm{QCD}}$ with that of the alternative hypothesis
that quark contact interactions are present in addition to QCD ${\cal
L}_{\rm{alt}}$:
\begin{eqnarray}
    {\cal R}_{LL} = \ln {\cal L}_{\rm{alt}} - \ln {\cal L}_{\rm{QCD}}.
\end{eqnarray}
The total likelihood is the product of the individual bin likelihoods, which
for $m_{jj}$ bin $i$ is
\begin{eqnarray}
\label{eq:lj}
  {\cal L}_i = {\cal P}(n_{{\rm tot},i}|\mu_{{\rm tot},i}){\cal B}(n_{{\rm
  in},i}|n_{{\rm tot},i},\rho_i),
\end{eqnarray}
where the first factor is the Poisson probability to observe $n_{{\rm
tot},i}$ events when expecting $\mu_{{\rm tot},i}$ and the second is the
binomial probability to observe $n_{\rm in,i}$ inner events given $n_{{\rm
tot},i}$ and a predicted probability to be inner of
$\rho_i~(\rho=R_\eta/(1+R_\eta))$.  Since the first factor in
Eq. (\ref{eq:lj}) contains no information on $R_{\eta}$, we remove it from
the statistical inference by conditioning the probabilities by the observed
values of $n_{{\rm tot},i}$~\cite{reid,dodders}.  We compare the value of
${\cal R}_{LL}$ in the data with distributions of the expected values for
both hypotheses, obtained from ensembles of pseudoexperiments, to either
claim the discovery of quark compositeness or set exclusion bounds on the
compositeness scale $\Lambda$ with the frequentist-inspired CL$_{\rm s}$
method~\cite{cls}.  This method provides protection against an exclusion
claim when the data have little sensitivity to the new physics.

We use the NLO prediction corrected for non-perturbative effects to describe
the shape of $R_{\eta}$ for the null hypothesis.  To minimize the effect of
potential discrepancies between the NLO prediction and actual QCD dijet
production, we include an overall offset of $R_{\eta}$ in the null
hypothesis.  This offset is determined with the data in the $m_{jj}$ range
between $490$ and $790$ GeV.  (The lower bound is chosen to avoid the region
where non-perturbative corrections are significant, and the upper bound is
chosen to avoid the signal region for compositeness.)  As noted above, the
data lie below the NLO prediction, yielding an offset of $\Delta R_{\eta} =
-0.050\pm0.021 ({\rm stat.})\pm0.039 ({\rm syst.})$.  Using ensembles of
simulated data, we determine that the probability ($p$-value) for observing
$\vert\Delta R_{\eta}\vert>0.050$, given the NLO prediction, is $0.29$.

\PYTHIA is used to describe $R_{\eta}$ for the alternative hypothesis.
We apply an $m_{jj}$-dependent correction that accounts for NLO
contributions to the QCD part of this prediction.  We do not apply this
correction, which is derived for $t$-channel QCD processes, to the contact
interaction part of the prediction because it is not physically motivated
and yields less conservative exclusion limits on $\Lambda$.  Since the
contact interaction model is not valid for $m_{jj}$ near the compositeness
scale, we exclude data above a $\Lambda$-dependent $m_{jj}$ threshold for
the testing of each $\Lambda$ value hypothesis.

\begin{figure}[hbtp!]
  \begin{center}
    \includegraphics[scale=0.39]{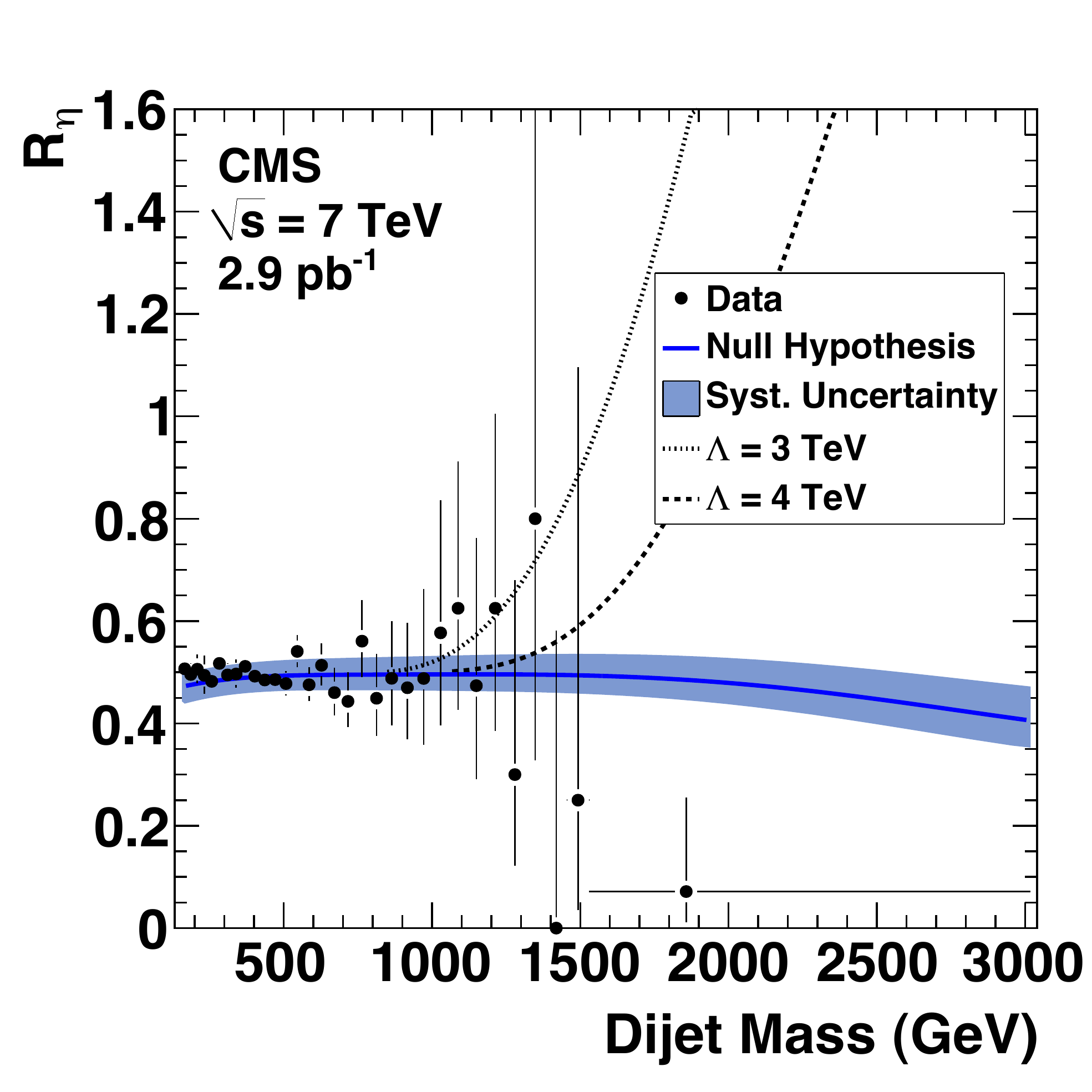}
    \caption{The observed dijet centrality ratio as a
    function of $m_{jj}$ compared with the null (QCD) hypothesis
    (solid line), including the total systematic uncertainty
    (band), and to hypotheses of quark contact interactions
    with $\Lambda = 3$ TeV (dotted line) and $4$ TeV
    (dashed line).}
\label{fig:data_np}
\end{center}
\end{figure}

In Table~\ref{tab:syst} we report the systematic uncertainties related to
the measurement of $R_{\eta}$ and the NLO QCD model.
\input{syst_tab}
The dominant source of uncertainty on the measurement is the 1\% uncertainty
in the relative jet energy scale (JES) between the inner and outer $\eta$
regions, which results in a 5--13\% uncertainty on $R_{\eta}$ depending on
$m_{jj}$.  This relative uncertainty has a much larger impact than a 10\%
uncertainty on the JES common to both regions.  For the QCD model, the
sources of uncertainty include the choice of scale and PDFs in the NLO
calculation and the non-perturbative corrections described above.  In
addition, we take the statistical uncertainty on the offset described above
and the difference between the \PYTHIA and NLO predictions as systematic
uncertainties related to our choice of model.  For the compositeness
hypothesis, $R_{\eta}$ increases steeply with $m_{jj}$, and the 10\%
uncertainty on the absolute JES dominates the uncertainty on the $\Lambda$
scale being probed.

Figure~\ref{fig:data_np} shows our data in comparison with the null
hypothesis.  Alternative hypotheses with contact interaction scales of
$\Lambda = 3$ and $4$ TeV are also shown.  In this figure, the data from the
15 sparsely populated $m_{jj}$ bins in the range 1530--3020 GeV are combined
into a single bin for presentation purposes.  The band indicates the total
systematic uncertainty, which is included in the ensembles of
pseudoexperiments with the method of Ref.~\cite{CousHigh}; i.e., the
uncertainties enter the ensembles as nuisance parameters that affect the
expected numbers of inner and outer events.

To quantify the agreement of the data with the SM expectation, we determine
the offset of the data with respect to the NLO model for the full $m_{jj}$
range, finding $-0.037\pm0.007 ({\rm stat.})\pm0.039 ({\rm syst.})$ with a
$p$-value of $0.34$.  Given this consistency of the data with the QCD
hypothesis, we determine $95\%$ CL limits on the contact interaction scale
$\Lambda$.

\begin{figure}[hbtp!]
\begin{center}
\includegraphics[scale=0.39]{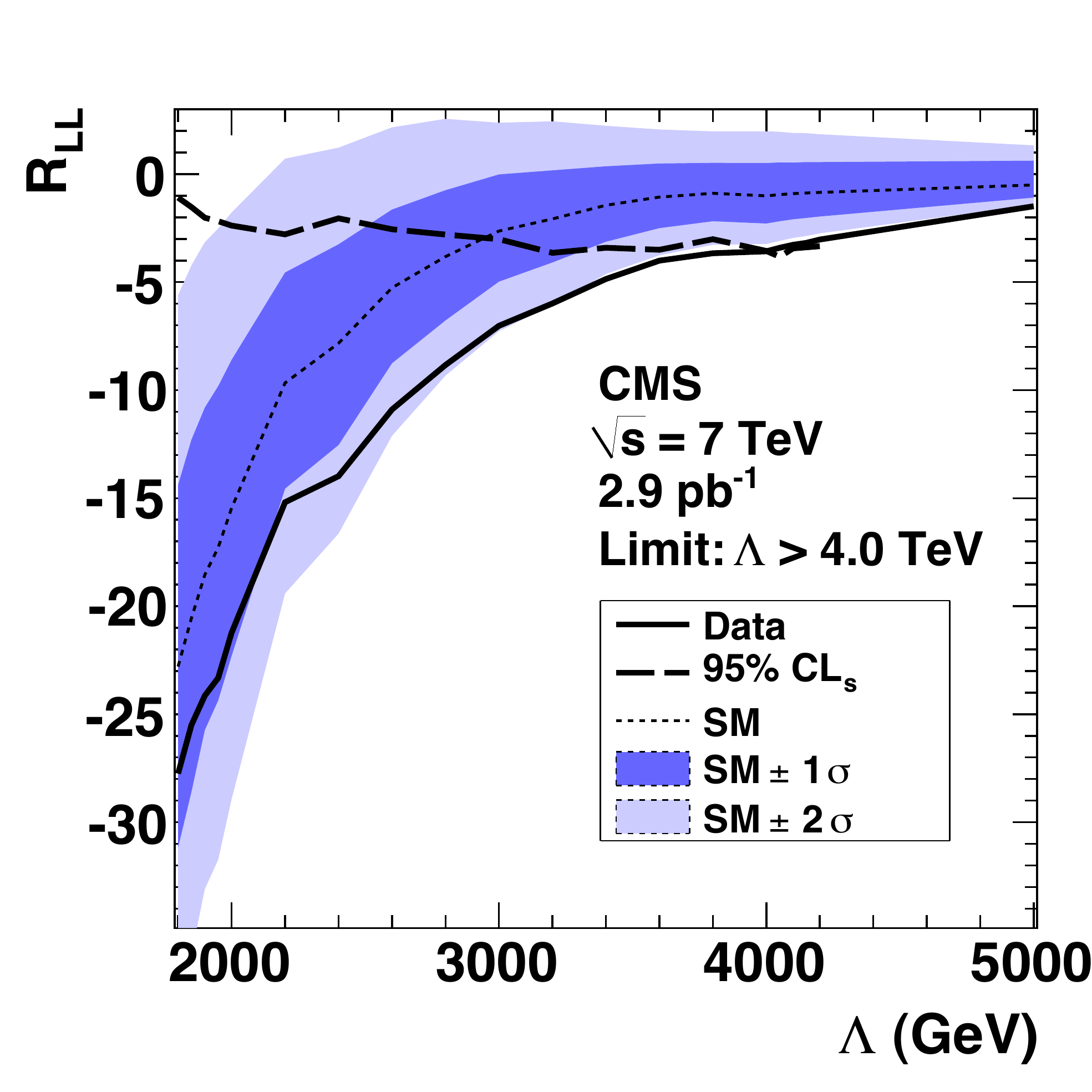}
\caption{Summary of the limit for the contact interaction
scale $\Lambda$.  We show ${\cal R}_{LL}$ versus $\Lambda$ for the data
  (solid line), the 95\% CL$_s$ (dashed line), and the SM expectation
  (dotted line) with 1$\sigma$ (dark) and 2$\sigma$ (light) bands.}
\label{fig:lim}
\end{center}
\end{figure}

We summarize the determination of the limit in Fig.~\ref{fig:lim}.  We show
${\cal R}_{LL}$ versus $\Lambda$ for the data and for the SM expectation
(with 1$\sigma$ and 2$\sigma$ bands) along with the highest value of ${\cal
R}_{LL}$ excluded at the 95\% CL with the CL$_s$ method.  The expected
exclusion region comprises those values of $\Lambda$ for which the
SM-expected ${\cal R}_{LL}$ (conditioned by the observed numbers of events
$n_{{\rm tot},i}$) is less than the 95\% CL$_s$ contour, and is seen to be
$\Lambda<2.9$ TeV.  The observed exclusion region comprises values for which
the measured ${\cal R}_{LL}$ is less than the $95\%$ CL$_s$ contour, and is
seen to be $\Lambda<4.0$ TeV.  The observed limit is higher than expected
because for $m_{jj}>1.4$ TeV the measured $R_\eta$ is lower than its
expectation under the SM.

In summary, we present a measurement of the dijet centrality ratio in 7 TeV
proton-proton collisions.  The dijet centrality ratio is found to exhibit
little dependence on the dijet invariant mass and to agree with the
expectation of the Standard Model.  We exclude quark compositeness described
by a contact interaction between left-handed quark fields at energy scales
of $\Lambda<4.0$ TeV at the 95\% CL.  This is the most stringent limit to
date.


We wish to congratulate our colleagues in the CERN accelerator departments
for the excellent performance of the LHC machine. We thank the technical and
administrative staff at CERN and other CMS institutes, and acknowledge
support from: FMSR (Austria); FNRS and FWO (Belgium); CNPq, CAPES, FAPERJ,
and FAPESP (Brazil); MES (Bulgaria); CERN; CAS, MoST, and NSFC (China);
COLCIENCIAS (Colombia); MSES (Croatia); RPF (Cyprus); Academy of Sciences
and NICPB (Estonia); Academy of Finland, ME, and HIP (Finland); CEA and
CNRS/IN2P3 (France); BMBF, DFG, and HGF (Germany); GSRT (Greece); OTKA and
NKTH (Hungary); DAE and DST (India); IPM (Iran); SFI (Ireland); INFN
(Italy); NRF and WCU (Korea); LAS (Lithuania); CINVESTAV, CONACYT, SEP, and
UASLP-FAI (Mexico); PAEC (Pakistan); SCSR (Poland); FCT (Portugal); JINR
(Armenia, Belarus, Georgia, Ukraine, Uzbekistan); MST and MAE (Russia); MSTD
(Serbia); MICINN and CPAN (Spain); Swiss Funding Agencies (Switzerland); NSC
(Taipei); TUBITAK and TAEK (Turkey); STFC (United Kingdom); DOE and NSF
(USA).

\bibliography{auto_generated}   

\cleardoublepage\appendix\section{The CMS Collaboration \label{app:collab}}\begin{sloppypar}\hyphenpenalty=5000\widowpenalty=500\clubpenalty=5000\input{EXO-10-002-authorlist.tex}\end{sloppypar}
\end{document}

%% file: ptdr-definitions.tex
%
%
%

\providecommand {\etal}{\mbox{et al.}\xspace} 
\providecommand {\ie}{\mbox{i.e.}\xspace}     
\providecommand {\eg}{\mbox{e.g.}\xspace}     
\providecommand {\etc}{\mbox{etc.}\xspace}     
\providecommand {\vs}{\mbox{\sl vs.}\xspace}      
\providecommand {\mdash}{\ensuremath{\mathrm{-}}} 

\providecommand {\Lone}{Level-1\xspace} 
\providecommand {\Ltwo}{Level-2\xspace}
\providecommand {\Lthree}{Level-3\xspace}

\providecommand{\ACERMC} {\textsc{AcerMC}\xspace}
\providecommand{\ALPGEN} {{\textsc{alpgen}}\xspace}
\providecommand{\CHARYBDIS} {{\textsc{charybdis}}\xspace}
\providecommand{\CMKIN} {\textsc{cmkin}\xspace}
\providecommand{\CMSIM} {{\textsc{cmsim}}\xspace}
\providecommand{\CMSSW} {{\textsc{cmssw}}\xspace}
\providecommand{\COBRA} {{\textsc{cobra}}\xspace}
\providecommand{\COCOA} {{\textsc{cocoa}}\xspace}
\providecommand{\COMPHEP} {\textsc{CompHEP}\xspace}
\providecommand{\EVTGEN} {{\textsc{evtgen}}\xspace}
\providecommand{\FAMOS} {{\textsc{famos}}\xspace}
\providecommand{\GARCON} {\textsc{garcon}\xspace}
\providecommand{\GARFIELD} {{\textsc{garfield}}\xspace}
\providecommand{\GEANE} {{\textsc{geane}}\xspace}
\providecommand{\GEANTfour} {{\textsc{geant4}}\xspace}
\providecommand{\GEANTthree} {{\textsc{geant3}}\xspace}
\providecommand{\GEANT} {{\textsc{geant}}\xspace}
\providecommand{\HDECAY} {\textsc{hdecay}\xspace}
\providecommand{\HERWIG} {{\textsc{herwig}}\xspace}
\providecommand{\HIGLU} {{\textsc{higlu}}\xspace}
\providecommand{\HIJING} {{\textsc{hijing}}\xspace}
\providecommand{\IGUANA} {\textsc{iguana}\xspace}
\providecommand{\ISAJET} {{\textsc{isajet}}\xspace}
\providecommand{\ISAPYTHIA} {{\textsc{isapythia}}\xspace}
\providecommand{\ISASUGRA} {{\textsc{isasugra}}\xspace}
\providecommand{\ISASUSY} {{\textsc{isasusy}}\xspace}
\providecommand{\ISAWIG} {{\textsc{isawig}}\xspace}
\providecommand{\MADGRAPH} {\textsc{MadGraph}\xspace}
\providecommand{\MCATNLO} {\textsc{mc@nlo}\xspace}
\providecommand{\MCFM} {\textsc{mcfm}\xspace}
\providecommand{\MILLEPEDE} {{\textsc{millepede}}\xspace}
\providecommand{\ORCA} {{\textsc{orca}}\xspace}
\providecommand{\OSCAR} {{\textsc{oscar}}\xspace}
\providecommand{\PHOTOS} {\textsc{photos}\xspace}
\providecommand{\PROSPINO} {\textsc{prospino}\xspace}
\providecommand{\PYTHIA} {{\textsc{pythia}}\xspace}
\providecommand{\SHERPA} {{\textsc{sherpa}}\xspace}
\providecommand{\TAUOLA} {\textsc{tauola}\xspace}
\providecommand{\TOPREX} {\textsc{TopReX}\xspace}
\providecommand{\XDAQ} {{\textsc{xdaq}}\xspace}

\providecommand {\DZERO}{D\O\xspace}     


\providecommand{\de}{\ensuremath{^\circ}}
\providecommand{\ten}[1]{\ensuremath{\times \text{10}^\text{#1}}}
\providecommand{\unit}[1]{\ensuremath{\text{\,#1}}\xspace}
\providecommand{\mum}{\ensuremath{\,\mu\text{m}}\xspace}
\providecommand{\micron}{\ensuremath{\,\mu\text{m}}\xspace}
\providecommand{\cm}{\ensuremath{\,\text{cm}}\xspace}
\providecommand{\mm}{\ensuremath{\,\text{mm}}\xspace}
\providecommand{\mus}{\ensuremath{\,\mu\text{s}}\xspace}
\providecommand{\keV}{\ensuremath{\,\text{ke\hspace{-.08em}V}}\xspace}
\providecommand{\MeV}{\ensuremath{\,\text{Me\hspace{-.08em}V}}\xspace}
\providecommand{\GeV}{\ensuremath{\,\text{Ge\hspace{-.08em}V}}\xspace}
\providecommand{\TeV}{\ensuremath{\,\text{Te\hspace{-.08em}V}}\xspace}
\providecommand{\PeV}{\ensuremath{\,\text{Pe\hspace{-.08em}V}}\xspace}
\providecommand{\keVc}{\ensuremath{{\,\text{ke\hspace{-.08em}V\hspace{-0.16em}/\hspace{-0.08em}}c}}\xspace}
\providecommand{\MeVc}{\ensuremath{{\,\text{Me\hspace{-.08em}V\hspace{-0.16em}/\hspace{-0.08em}}c}}\xspace}
\providecommand{\GeVc}{\ensuremath{{\,\text{Ge\hspace{-.08em}V\hspace{-0.16em}/\hspace{-0.08em}}c}}\xspace}
\providecommand{\TeVc}{\ensuremath{{\,\text{Te\hspace{-.08em}V\hspace{-0.16em}/\hspace{-0.08em}}c}}\xspace}
\providecommand{\keVcc}{\ensuremath{{\,\text{ke\hspace{-.08em}V\hspace{-0.16em}/\hspace{-0.08em}}c^\text{2}}}\xspace}
\providecommand{\MeVcc}{\ensuremath{{\,\text{Me\hspace{-.08em}V\hspace{-0.16em}/\hspace{-0.08em}}c^\text{2}}}\xspace}
\providecommand{\GeVcc}{\ensuremath{{\,\text{Ge\hspace{-.08em}V\hspace{-0.16em}/\hspace{-0.08em}}c^\text{2}}}\xspace}
\providecommand{\TeVcc}{\ensuremath{{\,\text{Te\hspace{-.08em}V\hspace{-0.16em}/\hspace{-0.08em}}c^\text{2}}}\xspace}

\providecommand{\pbinv} {\mbox{\ensuremath{\,\text{pb}^\text{$-$1}}}\xspace}
\providecommand{\fbinv} {\mbox{\ensuremath{\,\text{fb}^\text{$-$1}}}\xspace}
\providecommand{\nbinv} {\mbox{\ensuremath{\,\text{nb}^\text{$-$1}}}\xspace}
\providecommand{\percms}{\ensuremath{\,\text{cm}^\text{$-$2}\,\text{s}^\text{$-$1}}\xspace}
\providecommand{\lumi}{\ensuremath{\mathcal{L}}\xspace}
\providecommand{\Lumi}{\ensuremath{\mathcal{L}}\xspace}
%
\providecommand{\LvLow}  {\ensuremath{\mathcal{L}=\text{10}^\text{32}\,\text{cm}^\text{$-$2}\,\text{s}^\text{$-$1}}\xspace}
\providecommand{\LLow}   {\ensuremath{\mathcal{L}=\text{10}^\text{33}\,\text{cm}^\text{$-$2}\,\text{s}^\text{$-$1}}\xspace}
\providecommand{\lowlumi}{\ensuremath{\mathcal{L}=\text{2}\times \text{10}^\text{33}\,\text{cm}^\text{$-$2}\,\text{s}^\text{$-$1}}\xspace}
\providecommand{\LMed}   {\ensuremath{\mathcal{L}=\text{2}\times \text{10}^\text{33}\,\text{cm}^\text{$-$2}\,\text{s}^\text{$-$1}}\xspace}
\providecommand{\LHigh}  {\ensuremath{\mathcal{L}=\text{10}^\text{34}\,\text{cm}^\text{$-$2}\,\text{s}^\text{$-$1}}\xspace}
\providecommand{\hilumi} {\ensuremath{\mathcal{L}=\text{10}^\text{34}\,\text{cm}^\text{$-$2}\,\text{s}^\text{$-$1}}\xspace}


\providecommand{\zp}{\ensuremath{\mathrm{Z}^\prime}\xspace}


\providecommand{\kt}{\ensuremath{k_{\mathrm{T}}}\xspace}
\providecommand{\BC}{\ensuremath{\mathrm{B_{c}}}\xspace}
\providecommand{\bbarc}{\ensuremath{\mathrm{\overline{b}c}}\xspace}
\providecommand{\bbbar}{\ensuremath{\mathrm{b\overline{b}}}\xspace}
\providecommand{\ccbar}{\ensuremath{\mathrm{c\overline{c}}}\xspace}
\providecommand{\JPsi}{\ensuremath{\mathrm{J}\hspace{-.08em}/\hspace{-.14em}\psi}\xspace}
\providecommand{\bspsiphi}{\ensuremath{\mathrm{B_s} \to \JPsi\, \phi}\xspace}
\providecommand{\AFB}{\ensuremath{A_\text{FB}}\xspace}
\providecommand{\EE}{\ensuremath{\mathrm{e^+e^-}}\xspace}
\providecommand{\MM}{\ensuremath{\mu^+\mu^-}\xspace}
\providecommand{\TT}{\ensuremath{\tau^+\tau^-}\xspace}
\providecommand{\wangle}{\ensuremath{\sin^{2}\theta_{\text{eff}}^\text{lept}(M^2_\mathrm{Z})}\xspace}
\providecommand{\ttbar}{\ensuremath{\mathrm{t\overline{t}}}\xspace}
\providecommand{\stat}{\ensuremath{\,\text{(stat.)}}\xspace}
\providecommand{\syst}{\ensuremath{\,\text{(syst.)}}\xspace}

\providecommand{\HGG}{\ensuremath{\mathrm{H}\to\gamma\gamma}}
\providecommand{\gev}{\GeV}
\providecommand{\GAMJET}{\ensuremath{\gamma + \text{jet}}}
\providecommand{\PPTOJETS}{\ensuremath{\mathrm{pp}\to\text{jets}}}
\providecommand{\PPTOGG}{\ensuremath{\mathrm{pp}\to\gamma\gamma}}
\providecommand{\PPTOGAMJET}{\ensuremath{\mathrm{pp}\to\gamma + \mathrm{jet}}}
\providecommand{\MH}{\ensuremath{M_{\mathrm{H}}}}
\providecommand{\RNINE}{\ensuremath{R_\mathrm{9}}}
\providecommand{\DR}{\ensuremath{\Delta R}}


\providecommand{\PT}{\ensuremath{p_{\mathrm{T}}}\xspace}
\providecommand{\pt}{\ensuremath{p_{\mathrm{T}}}\xspace}
\providecommand{\ET}{\ensuremath{E_{\mathrm{T}}}\xspace}
\providecommand{\HT}{\ensuremath{H_{\mathrm{T}}}\xspace}
\providecommand{\et}{\ensuremath{E_{\mathrm{T}}}\xspace}
\providecommand{\Em}{\ensuremath{E\hspace{-0.6em}/}\xspace}
\providecommand{\Pm}{\ensuremath{p\hspace{-0.5em}/}\xspace}
\providecommand{\PTm}{\ensuremath{{p}_\mathrm{T}\hspace{-1.02em}/}\xspace}
\providecommand{\PTslash}{\ensuremath{{p}_\mathrm{T}\hspace{-1.02em}/}\xspace}
\providecommand{\ETm}{\ensuremath{E_{\mathrm{T}}^{\text{miss}}}\xspace}
\providecommand{\ETslash}{\ensuremath{E_{\mathrm{T}}\hspace{-1.1em}/}\xspace}
\providecommand{\MET}{\ensuremath{E_{\mathrm{T}}^{\text{miss}}}\xspace}
\providecommand{\ETmiss}{\ensuremath{E_{\mathrm{T}}^{\text{miss}}}\xspace}
\providecommand{\VEtmiss}{\ensuremath{{\vec E}_{\mathrm{T}}^{\text{miss}}}\xspace}

\providecommand{\dd}[2]{\ensuremath{\frac{\mathrm{d} #1}{\mathrm{d} #2}}}

%

\providecommand{\ga}{\ensuremath{\gtrsim}}
\providecommand{\la}{\ensuremath{\lesssim}}
\providecommand{\swsq}{\ensuremath{\sin^2\theta_\mathrm{W}}\xspace}
\providecommand{\cwsq}{\ensuremath{\cos^2\theta_\mathrm{W}}\xspace}
\providecommand{\tanb}{\ensuremath{\tan\beta}\xspace}
\providecommand{\tanbsq}{\ensuremath{\tan^{2}\beta}\xspace}
\providecommand{\sidb}{\ensuremath{\sin 2\beta}\xspace}
\providecommand{\alpS}{\ensuremath{\alpha_S}\xspace}
\providecommand{\alpt}{\ensuremath{\tilde{\alpha}}\xspace}

\providecommand{\QL}{\ensuremath{\mathrm{Q}_\mathrm{L}}\xspace}
\providecommand{\sQ}{\ensuremath{\tilde{\mathrm{Q}}}\xspace}
\providecommand{\sQL}{\ensuremath{\tilde{\mathrm{Q}}_\mathrm{L}}\xspace}
\providecommand{\ULC}{\ensuremath{\mathrm{U}_\mathrm{L}^\mathrm{C}}\xspace}
\providecommand{\sUC}{\ensuremath{\tilde{\mathrm{U}}^\mathrm{C}}\xspace}
\providecommand{\sULC}{\ensuremath{\tilde{\mathrm{U}}_\mathrm{L}^\mathrm{C}}\xspace}
\providecommand{\DLC}{\ensuremath{\mathrm{D}_\mathrm{L}^\mathrm{C}}\xspace}
\providecommand{\sDC}{\ensuremath{\tilde{\mathrm{D}}^\mathrm{C}}\xspace}
\providecommand{\sDLC}{\ensuremath{\tilde{\mathrm{D}}_\mathrm{L}^\mathrm{C}}\xspace}
\providecommand{\LL}{\ensuremath{\mathrm{L}_\mathrm{L}}\xspace}
\providecommand{\sL}{\ensuremath{\tilde{\mathrm{L}}}\xspace}
\providecommand{\sLL}{\ensuremath{\tilde{\mathrm{L}}_\mathrm{L}}\xspace}
\providecommand{\ELC}{\ensuremath{\mathrm{E}_\mathrm{L}^\mathrm{C}}\xspace}
\providecommand{\sEC}{\ensuremath{\tilde{\mathrm{E}}^\mathrm{C}}\xspace}
\providecommand{\sELC}{\ensuremath{\tilde{\mathrm{E}}_\mathrm{L}^\mathrm{C}}\xspace}
\providecommand{\sEL}{\ensuremath{\tilde{\mathrm{E}}_\mathrm{L}}\xspace}
\providecommand{\sER}{\ensuremath{\tilde{\mathrm{E}}_\mathrm{R}}\xspace}
\providecommand{\sFer}{\ensuremath{\tilde{\mathrm{f}}}\xspace}
\providecommand{\sQua}{\ensuremath{\tilde{\mathrm{q}}}\xspace}
\providecommand{\sUp}{\ensuremath{\tilde{\mathrm{u}}}\xspace}
\providecommand{\suL}{\ensuremath{\tilde{\mathrm{u}}_\mathrm{L}}\xspace}
\providecommand{\suR}{\ensuremath{\tilde{\mathrm{u}}_\mathrm{R}}\xspace}
\providecommand{\sDw}{\ensuremath{\tilde{\mathrm{d}}}\xspace}
\providecommand{\sdL}{\ensuremath{\tilde{\mathrm{d}}_\mathrm{L}}\xspace}
\providecommand{\sdR}{\ensuremath{\tilde{\mathrm{d}}_\mathrm{R}}\xspace}
\providecommand{\sTop}{\ensuremath{\tilde{\mathrm{t}}}\xspace}
\providecommand{\stL}{\ensuremath{\tilde{\mathrm{t}}_\mathrm{L}}\xspace}
\providecommand{\stR}{\ensuremath{\tilde{\mathrm{t}}_\mathrm{R}}\xspace}
\providecommand{\stone}{\ensuremath{\tilde{\mathrm{t}}_1}\xspace}
\providecommand{\sttwo}{\ensuremath{\tilde{\mathrm{t}}_2}\xspace}
\providecommand{\sBot}{\ensuremath{\tilde{\mathrm{b}}}\xspace}
\providecommand{\sbL}{\ensuremath{\tilde{\mathrm{b}}_\mathrm{L}}\xspace}
\providecommand{\sbR}{\ensuremath{\tilde{\mathrm{b}}_\mathrm{R}}\xspace}
\providecommand{\sbone}{\ensuremath{\tilde{\mathrm{b}}_1}\xspace}
\providecommand{\sbtwo}{\ensuremath{\tilde{\mathrm{b}}_2}\xspace}
\providecommand{\sLep}{\ensuremath{\tilde{\mathrm{l}}}\xspace}
\providecommand{\sLepC}{\ensuremath{\tilde{\mathrm{l}}^\mathrm{C}}\xspace}
\providecommand{\sEl}{\ensuremath{\tilde{\mathrm{e}}}\xspace}
\providecommand{\sElC}{\ensuremath{\tilde{\mathrm{e}}^\mathrm{C}}\xspace}
\providecommand{\seL}{\ensuremath{\tilde{\mathrm{e}}_\mathrm{L}}\xspace}
\providecommand{\seR}{\ensuremath{\tilde{\mathrm{e}}_\mathrm{R}}\xspace}
\providecommand{\snL}{\ensuremath{\tilde{\nu}_L}\xspace}
\providecommand{\sMu}{\ensuremath{\tilde{\mu}}\xspace}
\providecommand{\sNu}{\ensuremath{\tilde{\nu}}\xspace}
\providecommand{\sTau}{\ensuremath{\tilde{\tau}}\xspace}
\providecommand{\Glu}{\ensuremath{\mathrm{g}}\xspace}
\providecommand{\sGlu}{\ensuremath{\tilde{\mathrm{g}}}\xspace}
\providecommand{\Wpm}{\ensuremath{\mathrm{W}^{\pm}}\xspace}
\providecommand{\sWpm}{\ensuremath{\tilde{\mathrm{W}}^{\pm}}\xspace}
\providecommand{\Wz}{\ensuremath{\mathrm{W}^{0}}\xspace}
\providecommand{\sWz}{\ensuremath{\tilde{\mathrm{W}}^{0}}\xspace}
\providecommand{\sWino}{\ensuremath{\tilde{\mathrm{W}}}\xspace}
\providecommand{\Bz}{\ensuremath{\mathrm{B}^{0}}\xspace}
\providecommand{\sBz}{\ensuremath{\tilde{\mathrm{B}}^{0}}\xspace}
\providecommand{\sBino}{\ensuremath{\tilde{\mathrm{B}}}\xspace}
\providecommand{\Zz}{\ensuremath{\mathrm{Z}^{0}}\xspace}
\providecommand{\sZino}{\ensuremath{\tilde{\mathrm{Z}}^{0}}\xspace}
\providecommand{\sGam}{\ensuremath{\tilde{\gamma}}\xspace}
\providecommand{\chiz}{\ensuremath{\tilde{\chi}^{0}}\xspace}
\providecommand{\chip}{\ensuremath{\tilde{\chi}^{+}}\xspace}
\providecommand{\chim}{\ensuremath{\tilde{\chi}^{-}}\xspace}
\providecommand{\chipm}{\ensuremath{\tilde{\chi}^{\pm}}\xspace}
\providecommand{\Hone}{\ensuremath{\mathrm{H}_\mathrm{d}}\xspace}
\providecommand{\sHone}{\ensuremath{\tilde{\mathrm{H}}_\mathrm{d}}\xspace}
\providecommand{\Htwo}{\ensuremath{\mathrm{H}_\mathrm{u}}\xspace}
\providecommand{\sHtwo}{\ensuremath{\tilde{\mathrm{H}}_\mathrm{u}}\xspace}
\providecommand{\sHig}{\ensuremath{\tilde{\mathrm{H}}}\xspace}
\providecommand{\sHa}{\ensuremath{\tilde{\mathrm{H}}_\mathrm{a}}\xspace}
\providecommand{\sHb}{\ensuremath{\tilde{\mathrm{H}}_\mathrm{b}}\xspace}
\providecommand{\sHpm}{\ensuremath{\tilde{\mathrm{H}}^{\pm}}\xspace}
\providecommand{\hz}{\ensuremath{\mathrm{h}^{0}}\xspace}
\providecommand{\Hz}{\ensuremath{\mathrm{H}^{0}}\xspace}
\providecommand{\Az}{\ensuremath{\mathrm{A}^{0}}\xspace}
\providecommand{\Hpm}{\ensuremath{\mathrm{H}^{\pm}}\xspace}
\providecommand{\sGra}{\ensuremath{\tilde{\mathrm{G}}}\xspace}
\providecommand{\mtil}{\ensuremath{\tilde{m}}\xspace}
\providecommand{\rpv}{\ensuremath{\rlap{\kern.2em/}R}\xspace}
\providecommand{\LLE}{\ensuremath{LL\bar{E}}\xspace}
\providecommand{\LQD}{\ensuremath{LQ\bar{D}}\xspace}
\providecommand{\UDD}{\ensuremath{\overline{UDD}}\xspace}
\providecommand{\Lam}{\ensuremath{\lambda}\xspace}
\providecommand{\Lamp}{\ensuremath{\lambda'}\xspace}
\providecommand{\Lampp}{\ensuremath{\lambda''}\xspace}
\providecommand{\spinbd}[2]{\ensuremath{\bar{#1}_{\dot{#2}}}\xspace}

\providecommand{\MD}{\ensuremath{{M_\mathrm{D}}}\xspace}
\providecommand{\Mpl}{\ensuremath{{M_\mathrm{Pl}}}\xspace}
\providecommand{\Rinv} {\ensuremath{{R}^{-1}}\xspace} 

%% file: syst_tab.tex
\begin{table}[!htbp]

\caption{Systematic uncertainties on $R_\eta$ related to the measurement of
  $R_\eta$ (detector uncertainties) and to the QCD model (model
  uncertainties). For each source of uncertainty, we show the range of
  values over the entire $m_{jj}$ range and at a representative point in the
  signal region.}

\label{tab:syst}
\begin{center}
\begin{tabular}{lcc}
\hline\hline
  Source     & Full Range  &  $m_{jj}=1.6$ TeV \\ 
\hline

Detector uncertainty\\
~~Relative JES                 & $0.02 \mbox{-} 0.05$ & $0.032$\\ 
~~Absolute JES               & $0.00 \mbox{-} 0.03$ & $0.003$\\
~~Jet Energy Resolution & $0.003$                      & $0.003$ \\
~~Other                         & $0.01$                       & $0.010$ \\
~~\bf{Total Detector}& \boldmath{$0.02 \mbox{-} 0.05 $} & \boldmath{$0.034$} \\
\hline
Model uncertainty\\
~~PYTHIA6$-$NLO & $0.00 \mbox{-} 0.05$ & $0.032$\\
~~Offset              & $0.021$ & $0.021$ \\
~~Scale               & $^{+(0.01 \mbox{-} 0.05)}_{-(0.01 \mbox{-} 0.02)}$ & $^{+0.029}_{-0.011}$\\
~~PDF                 & $^{+(0.002 \mbox{-} 0.004)}_{-(0.002 \mbox{-} 0.007)}$ & $^{+0.002}_{-0.003}$\\
~~MC Statistics       & $0.005$ & $0.005$ \\
~~Non-pert. Corr.     & $0.002 \mbox{-} 0.014$ & $0.002$\\
~~\bf{Total Model}    & \boldmath{$^{+(0.02 \mbox{-} 0.07)}_{-(0.01 \mbox{-} 0.05)}$} & \boldmath{$^{+0.044}_{-0.034}$} \\
\hline
\bf{Total} & \boldmath{$^{+(0.03 \mbox{-} 0.09)}_{-(0.03 \mbox{-} 0.08)}$} & \boldmath{$^{+0.055}_{-0.048}$} \\
\hline\hline
\end{tabular}
\end{center}
\end{table}

%% file: EXO-10-002-authorlist.tex
\textbf{Yerevan Physics Institute,  Yerevan,  Armenia}\\*[0pt]
V.~Khachatryan, A.M.~Sirunyan, A.~Tumasyan
\vskip\cmsinstskip
\textbf{Institut f\"{u}r Hochenergiephysik der OeAW,  Wien,  Austria}\\*[0pt]
W.~Adam, T.~Bergauer, M.~Dragicevic, J.~Er\"{o}, C.~Fabjan, M.~Friedl, R.~Fr\"{u}hwirth, V.M.~Ghete, J.~Hammer\cmsAuthorMark{1}, S.~H\"{a}nsel, C.~Hartl, M.~Hoch, N.~H\"{o}rmann, J.~Hrubec, M.~Jeitler, G.~Kasieczka, W.~Kiesenhofer, M.~Krammer, D.~Liko, I.~Mikulec, M.~Pernicka, H.~Rohringer, R.~Sch\"{o}fbeck, J.~Strauss, A.~Taurok, F.~Teischinger, W.~Waltenberger, G.~Walzel, E.~Widl, C.-E.~Wulz
% ...
\vskip\cmsinstskip
% ...
1:~~Also at CERN, European Organization for Nuclear Research, Geneva, Switzerland\\
% ...
